\documentclass[a4paper,11pt]{article}

\usepackage{amsmath,amssymb}
\usepackage[dvipdfmx]{graphicx}
\usepackage[dvipsnames]{xcolor}
\numberwithin{equation}{section}
\usepackage[colorinlistoftodos]{todonotes}
\usepackage{cite}
\usepackage{color}
\usepackage{mathrsfs}
\usepackage[top=2.828cm,bottom=2.828cm, left=2.5cm,right=2.5cm]{geometry}
\usepackage{caption}
\usepackage{subcaption}
\definecolor{lgray}{gray}{0.35}
\usepackage[colorlinks=true,urlcolor=Gray,linkcolor=lgray,
citecolor=Gray,breaklinks=true,linktocpage=true,pdfpagelabels=true,
hypertexnames=true,plainpages=false,naturalnames=false]{hyperref}
\newcommand{\be}{\begin{equation}}
\newcommand{\ee}{\end{equation}}
\newcommand{\bea}{\begin{eqnarray}}
\newcommand{\eea}{\end{eqnarray}}

\newcommand{\de}{{\rm d}}
\newcommand{\K}{{\boldsymbol k}}
\newcommand{\x}{{\boldsymbol x}}
\newcommand{\q}{{\boldsymbol q}}
\newcommand{\y}{{\boldsymbol y}}

\begin{document}

\setlength\arraycolsep{2pt}

\renewcommand{\theequation}{\arabic{section}.\arabic{equation}}
\setcounter{page}{1}

\begin{titlepage}

\begin{center}

{\Huge Note on initial conditions for small-f{i}eld inf{l}ation}

\vskip 2.0cm

{\large Ignatios Antoniadis$^\text{\scriptsize1,2}$, Auttakit Chatrabhuti$^\text{\scriptsize3}$, Hiroshi Isono$^\text{\scriptsize3}$ and Spyros Sypsas$^\text{\scriptsize3,4}$}
\vskip .5cm
$^\text{\scriptsize1}${\it Laboratoire de Physique Th\'eorique et Hautes Energies (LPTHE),
UMR 7589, Sorbonne Universit\'e et CNRS,
4 place Jussieu, 75252 Paris Cedex 05, France} \\
$^\text{\scriptsize2}${\it Albert Einstein Center, Institute for Theoretical Physics, University of Bern,
Sidlerstrasse 5, 3012 Bern, Switzerland} \\
$^\text{\scriptsize3}${\it Department of Physics, Faculty of Science, Chulalongkorn University, Phayathai Rd., Bangkok 10330, Thailand}\\
$^\text{\scriptsize4}${\it National Astronomical Research Institute of Thailand, Don Kaeo, Mae Rim, Chiang Mai 50180, Thailand}

\vskip 2.5cm

\end{center}

\begin{abstract} 
We show that initial conditions for small-field inflation can be determined quantum mechanically by introducing a suitable flattened region in the scalar potential. The inflaton is then driven towards the slow-roll attractor solution exponentially fast, desensitising inflation from the initial velocity and partially evading the so-called overshoot problem. We give an explicit
example in the context of hilltop inflation by introducing an ultra slow-roll plateau around the
maximum of the potential and analyse its effect on the phase-space trajectories. 
\end{abstract}

\end{titlepage}

\section{Introduction}
Inflation is the dominant solution to the shortcomings of the standard cosmological model~\cite{Guth:1980zm, Linde:1981mu, Albrecht:1982wi, Starobinsky:1980te, Mukhanov:1981xt}, providing us with a theoretical explanation of the observed Gaussian, adiabatic and scale invariant cosmic microwave background (CMB) temperature fluctuations. 
A fundamental aspect of the inflationary dynamics is the problem of initial conditions~\cite{GOLDWIRTH1992223,Brandenberger:2016uzh}: \emph{how possible is it for inflation to begin in a potential landscape?}
Among the questions regarding the initial state of the Universe lies the issue of fine-tuning the phase-space patch from which a successful slow-roll trajectory begins. If this patch is too small compared to the available phase-space then the model under consideration is plagued by the so-called overshoot problem~\cite{Brustein:1992nk}.
 
Inflationary models can be roughly divided into two broad categories of large-field and small-field depending on the distance in field space that the inflaton has to trace during its slow-rolling from an initial value, associated with the period when the modes responsible for the CMB anisotropies exit the Hubble horizon, to a final value signalling the onset of reheating; that is, along the $60$ e-folds required to solve the horizon and flatness problems (or a bit less, up to $50$, when the scale of inflation is at lower energies, up to TeV). The fine-tuning of the initial phase-space patch is then easy to treat in large-field models like chaotic inflation: no matter how rare a causally connected flat patch is, if there is one --- and there must be one, statistically speaking --- then it will inflate and result in our observable Universe.

Attracting as they might be from this perspective, large-field models face other issues. For example, the framework used to address inflation, general relativity (GR) itself, should be viewed as an effective field theory, with the Planck mass $M_{\rm Pl}\simeq 2.4\times 10^{18}$ GeV the ultraviolet (UV) cutoff marking the quantum gravity regime. The fact that the field has to trace a trans-Planckian distance has thus implications on the validity of inflation as an effective process. Recently, this has led to a quite active discussion on the possibility of realising inflation in a quantum gravity framework (swampland program, for reviews see~\cite{Brennan:2017rbf, Palti:2019pca}). 

On the other hand, in small-field inflation, where the coarse-graining of spacetime into causally connected patches is less refined, it becomes less probable for the right initial condition to be met and consequently for inflation to begin~\cite{Brandenberger:2016uzh}. A realisation of this is the overshoot problem in hilltop inflation (or in general, inflection point inflation): if the kinetic energy is too large or the initial inflaton value is slightly displaced from the maximum, then inflation lasts for much less than the required $60$ e-folds.

In this note, we focus on small-field models and consider a flattened region of the potential around the point when CMB scales exit the horizon.  Along this ultra flat patch, the classical dynamics is driven by the friction due to the expanding spacetime, which eventually forces the field to a halt, no matter how large the kinetic energy is. This shares many features with the plateau models studied in~\cite{Martin:2013nzq,Chowdhury:2019otk}. However, here we extend the idea and consider a plateau region as a generic \emph{module} of any small-field model. We show that as the velocity diminishes, at some point the contribution of the long wavelength quantum fluctuations becomes dominant and one can set initial conditions quantum mechanically\footnote{See also Ref.~\cite{Brahma:2020cpy} for another way to set quantum initial conditions in hilltop scenarios via tunneling of the initial field value to a position in the neighbourhood of the maximum.}. This is an initial condition in the sense that, given a long enough plateau, all the previous dynamics is rendered irrelevant, since the field eventually becomes dominated by the  fluctuations. 

Restricting the initial field value to be close to the end of this plateau allows us to compute the root-mean-square values of the fluctuations via the standard quantisation procedure around de Sitter (dS) spacetime, which uniquely fixes the initial velocity. Considering the phase-space from this point onwards, there is no overshoot. However, depending on the ratio of the kinetic to potential energy, the field might have travelled a trans-Planckian distance until it diminishes enough so that the long wave fluctuations take over, which is where we set initial conditions for slow-roll inflation. By adding such a plateau module, either far away from the end of inflation or on the hilltop, we thus turn a small-field model into a large-field one, trading the overshoot problem with a trans-Planckian displacement. There are several mechanisms leading to asymptotic plateaux or flattened hilltops, which we briefly outline.
Alternatively, in hilltop scenarios, starting at the maximum of the potential with classically zero speed may be motivated by symmetry, if at high temperatures this point is a symmetric minimum which became a maximum after spontaneous symmetry breaking occurring in a different field direction when temperature cooled down. In this case, our calculation of initial conditions is still valid.

Before passing to the main subject let us first set the notation that we will use in the rest of the paper by discussing a few general points.
\paragraph{Preliminaries/Notation}
The background FLRW spacetime, sourced by the homogeneous inflaton field $\bar{\phi}(t)$, is described by a line element given by
\begin{align} \label{frw}
\de s^2=a(\tau)^2(-\de \tau^2+\de \x^2),
\end{align}
where $\tau\in(-\infty,0)$ is the conformal time, defined as $a(\tau)\de \tau = \de t$. The Hubble expansion rate is given by 
\be 
H = \frac{\de \ln a}{\de t}, 
\ee
and the e-fold number is defined as
\be 
N\equiv \ln a = \int \de t\; H \simeq Ht,
\ee
the last equality being exact for a constant Hubble rate, in which case the metric~\eqref{frw} describes the flat slicing of de Sitter space, with 
\be 
a(\tau)=(-H\tau)^{-1}.
\ee
Depending on the situation, and since there is a one-to-one map among them, we will use the cosmic ($t$), conformal ($\tau$) and e-fold number ($N$) as time variables interchangeably (where there is no confusion). We denote cosmic time derivatives with overdots $(\dot{\phantom{a}})$ and derivatives with respect to e-fold with a prime $(')$. Moreover, when referring to specific points in field-space --- corresponding to specific moments --- we will further replace the time variable with an index; to summarise:
$$
\bar{\phi}(t_i) \equiv \bar{\phi}(\tau_i) \equiv \bar{\phi}(N_i) \equiv  \bar{\phi}_i .
$$
\paragraph{Organisation of the paper} We begin in Sec.~\ref{sec:overshoot} by describing the problem of initial conditions in small-field hilltop inflation and discuss our proposal to compute the initial conditions quantum mechanically around dS space (subsection~\ref{sec:sym}). A possible implementation is by introducing an extended flat region around the maximum that could be a consequence of effective interactions of the inflaton with other fields present in the underlying fundamental theory of gravity at high energies (subsection~\ref{sec:dec-ph}). In Sec.~\ref{sec:ICs}, we study the implications of such a flat region resulting in an ultra slow-roll phase (subsection~\ref{sec:bcgd}) and perform a quantum computation of the initial conditions for inflation at the horizon exit that should be somewhat earlier than the beginning of the usual classical slow-roll trajectory (subsection~\ref{sec:fluct}). In Sec.~\ref{sec:hilltop}, we present an explicit example in the hilltop framework and work out its implications and predictions. Finally, we conclude in Sec.~\ref{sec:conc}.

\section{Overshoot in small-field models}
\label{sec:overshoot}
In order to exemplify the problem, we will consider a well-studied small-field model, hilltop inflation~\cite{Boubekeur:2005zm}. In this scenario the potential has a maximum $V_{\rm m}$, located at $\bar{\phi}=\bar{\phi}_{\rm m}$, around which we may expand the potential as
\be
V\left( \bar{\phi}_{\rm m} + \Delta\bar{\phi} \right) = V_{\rm m}\left(1 - \frac{|\eta|}2\frac{\Delta\bar{\phi}^2}{M_{\rm Pl}^2} + \dots \right).
\label{scalar_potential1}
\ee

Focusing on the initial dynamics of the field, we may neglect the higher order terms in the potential. We further assume that $|\eta|\ll1$ so that the field slowly rolls down the potential for at least 60 e-folds. The dynamics is thus specified, in cosmic time, by
\be
\ddot{\bar{\phi}} + 3H\dot{ \bar{\phi}} - |\eta|  \frac{V_{\rm m}}{M_{\rm Pl}^2}\bar{\phi} = 0,
\label{eom}
\ee
and
\be
H^2=\frac{1}{6M_{\rm Pl}^2} \left(\dot {\bar{\phi}}^2 + 2 V\right).
\label{fried}
\ee

Being second order in derivatives, the system classically requires two independent initial conditions $(\bar{\phi}_{\rm i},\dot {\bar{\phi}}_{\rm i})$ to specify a solution. However, since the potential is such that the field is slowly rolling and the expansion of spacetime induces friction, via the term $3H\dot{ \bar{\phi}}$, the initial velocity is irrelevant, or put in other words, the system flows to an attractor solution for which $|\ddot{\bar{\phi}}|\ll|H\dot{\bar{\phi}}|$ and $H^2\propto V$. Nevertheless, if $|\dot {\bar{\phi}}_{\rm i}|$ is large and/or $\bar{\phi}_{\rm i}$ is slightly displaced from the maximum, the system might never reach this attractor within the $60$ observable e-folds. This is the so-called overshoot problem~\cite{Brustein:1992nk} which plagues small-field inflation models. For instance, a numerical evaluation of~\eqref{eom} and~\eqref{fried} with $(\bar\phi_{\rm i},\dot{\bar\phi}_{\rm i})=(\bar\phi_{\rm m}, 10^{-4}M_{\rm Pl}^2)$ reveals that for this case the field spends only $25$ e-folds in the slow-roll phase, which is much less than the 60 e-folds required for successful inflation.

\subsection{Invoking symmetries} \label{sec:sym}
One way to evade this is to assume that the aforementioned symmetry at high energies classically selects the initial condition $(\bar{\phi}_{\rm i},\dot {\bar{\phi}}_{\rm i})=(\bar{\phi}_{\rm m},0)$, i.e. the field is lying at the symmetric point with vanishing initial velocity. This configuration leads to a dS background with $\bar{\phi}(N)=\bar{\phi}_{\rm m}$ and $3M_{\rm Pl}^2H^2=V_{\rm m}$.

During inflation the homogeneous field undergoes quantum perturbations resulting in a weakly inhomogeneous field
\be \label{split}
\phi(t,\x)= \bar\phi(t)+\delta\phi(t,\x),
\ee
which sources the observed CMB temperature anisotropies and eventually  the large-scale structure of the Universe. In the case under consideration, the long wavelength Fourier modes of these fluctuations will spontaneously destabilise the initial background configuration. Since the only energy scale in the problem is the dS radius, set by $H$, the average position/velocity of the long modes will induce a quantum initial condition for the background 
\be \label{H-ic}
(\bar{\phi}_{\rm i},\dot {\bar{\phi}}_{\rm i})\simeq (H,H^2),
\ee
where we neglected numerical factors\footnote{The exact values, which we compute in the next section, are given in terms of the dS temperature $T_{\rm dS}=H/2\pi$. Furthermore, there is a time dependence in the both $\bar{\phi}_{\rm i}$ and $\dot{\bar{\phi}}_{\rm i}$ stemming from well-known properties of field theory on dS space~\cite{Vilenkin:1982wt,Linde:1982uu,Starobinsky:1982ee,Allen:1985ux,Ratra:1984yq,Ford:1984hs,Starobinsky:1986fx,Antoniadis:1986sb,Habib:1992ci}.}. Since $V_{\rm m}\gg H^4$, the Friedmann equation reads $3M_{\rm Pl}^2H^2=V_{\rm m}$, and we can solve~\eqref{eom} to obtain, to first order\footnote{The parameter $|\eta|$ here is fixed by the spectral index of the curvature power spectrum measured by Planck to $|\eta|\sim 0.02$.} in $\eta$,
\be
\bar{\phi}(N) = \bar{\phi}_{\rm m} + \frac{H}{9} \left( 3+(3N-2)|\eta| \right) \left(1- e^{-3N}\right).
\ee
Now $\ddot{\bar{\phi}}/(H\dot{\bar{\phi}})\sim e^{-3N}$, such that after a couple of e-folds we have $\ddot{\bar{\phi}}/(H\dot{\bar{\phi}})\ll1$ and the attractor is reached. Solving the system~\eqref{eom} and~\eqref{fried} numerically we can verify that indeed in this case the field spends around $60$ e-folds slowly rolling before it reaches the minimum of the potential at zero energy, where the neglected terms become relevant stabilising the potential and triggering a reheating phase.

However, in the most general situation where inflation is embedded in some ultraviolet framework, there might be more energy scales, and hence, from a bottom-up point of view, the initial velocity should be considered as a free parameter setting the UV cutoff of the effective field theory description, with an upper bound given by the Planck scale. As mentioned in the previous paragraph, this brings us back to the overshoot problem.
\subsection{Initial decelerating phase}
\label{sec:dec-ph}
In order to ameliorate this problem we will consider a situation where the field begins in a decelerating phase
such that whatever the initial velocity is, it (classically) gets dissipated away rendering any previous dynamics irrelevant. At this point, when the background field lies nearly frozen at some position, we can consider the long wavelength modes of the quantum fluctuation as setting initial conditions for the later evolution. There are a number of ways to slow down the inflaton field like particle production~\cite{Kofman:2004yc, Green:2009ds}, or the presence of a nearly flat region of the potential, which can also be attributed to interactions with heavy degrees of freedom~\cite{Dong:2010in} (see also~\cite{Antoniadis:2019dpm} for recent developments in a supergravity framework and~\cite{Geng:2019phi} for a string theoretic interpretation). 

Here, we will consider the latter possibility: an extended ultra flat region around the point where we set initial conditions. 
A simple example of this sort, due to the presence\footnote{As pointed out in~\cite{Dong:2010in} (see also~\cite{Geng:2019phi}), a heavy field coupled to the inflaton does not guarantee the flatness of the potential; in a realistic case, details of the UV physics, such as, e.g., the compactification mechanism, might spoil this feature.} of a heavy field, is given in Ref.~\cite{Dong:2010in} (see also~\cite{Pedro:2019klo}): consider a light inflaton $\varphi$ and a heavy scalar $\chi$ interacting via 
$
V(\varphi,\chi)=g^2\varphi^2\chi^2+M_\chi^2(\chi-\chi_0)^2.
$
Assuming $M_\chi\gg H$, minimising along the heavy direction and considering the large-field region $\varphi\gg M_\chi/g$, results in 
\be  \label{v-example} 
V(\varphi)\simeq M_\chi^2\chi_0^2 \left(1 - \frac{M_\chi^2}{g^2\varphi^2}\right),
\ee
which is quite flat around where one sets the initial condition, far (at least 60 e-folds) from the point marking the end of inflation. 
Such potentials arise typically in brane inflation~\cite{Kachru:2003sx, Kallosh:2018zsi}. 

A similar effect of flattening the potential far away from the minimum can be also obtained in the Palatini formulation of GR where the metric and the connection are treated as independent variables. Although pure gravity is equivalent in both formulations, this is not the case when scalar fields are present with non-minimal couplings, as well as in the presence of an $R^2$ term (see~\cite{Sotiriou:2008rp} for a review). In this case, it turns out that the Palatini formulation of a model described by a potential $V(\phi)$ leads to the creation of a plateau for the Einstein-frame potential $\bar{V}(\phi)=V(\phi)/\left(1+4\alpha V(\phi)/M_{\rm Pl}^4\right)$ at large field values where $\alpha$ is the $R^2$ coupling, providing new possibilities for slow-roll inflation~\cite{Antoniadis:2018ywb, Enckell:2018hmo}. Indeed whatever the origin of the plateau is, such models have been shown to be free of fine-tuned initial conditions~\cite{Chowdhury:2019otk,Tenkanen:2020cvw}.

Another mechanism leading to the flattening of the potential around extrema --- hence, appropriate for hilltop scenarios --- is the running of the inflaton mass under renormalisation group flow~\cite{Stewart:1996ey, Stewart:1997wg}. This induces a dependence of the curvature of the potential on the field value, and according to the UV embedding of inflation, it can be made small enough so that the dynamics around the maximum exhibits a decelerating phase. Finally, the higher order terms neglected in \eqref{scalar_potential1} may have a similar effect.

From a landscape perspective~\cite{Susskind:2003kw}, all of the above cases can be thought of as specific points in the moduli space of the parent theory, wherein inflation operates, leading to randomly distributed ultra flat regions of some multifield potential arising from e.g. string compactifications. Initial conditions can then be understood as quantum tunneling from such initial patches to the slow-roll regime~\cite{Kawasaki:2015mvm,Masoumi:2016eag,Masoumi:2017gmh}.

Our purpose here is not to study a particular realisation but to argue that a decelerating phase as a module of an inflationary model desensitises inflation from UV completions, in the sense that the initial kinetic energy is dynamically set by the Hubble scale. 
Perturbation theory along such extended flat regions in the inflaton potential has been considered in the literature for several purposes, among which are the production of primordial black holes~\cite{Kohri:2007qn, Garcia-Bellido:2017mdw, Motohashi:2017kbs, Germani:2017bcs}, if this plateau lies close to the end of inflation, as well as the study of ultra slow-roll (USR) inflation~\cite{Tsamis:2003px, Kinney:2005vj}, whose non-attractor nature serves as an interesting single-field counterexample~\cite{Namjoo:2012aa} to the local non-Gaussianity consistency condition~\cite{Maldacena:2002vr}. 
Here, we will use the friction-driven USR motion in order to dynamically set quantum initial conditions for small-field models given by Eq.~\eqref{H-ic}. 
We thus trade the problem of a fine-tuned initial phase-space patch with the presence of a long and flat enough region on the hilltop; in the latter case though, the flatness of the potential might be justified from a UV point of view. 

\section{Quantum initial conditions on ultra slow-roll plateaux}
\label{sec:ICs}
\subsection{Background dynamics} \label{sec:bcgd}
We may define the USR motion along the ultra flat valley by the following condition:
\be  \label{usr-def}
\left| \ddot{\bar{\phi}}(t) \right| \gg  \left|\frac{\de V}{\de \bar{\phi}}\right|.
\ee
The width of this region may be determined by the breakdown of this condition at $\bar{\phi}=\bar{\phi}_{\rm c}$:
\be  \label{usr-sr}
\ddot{\bar{\phi}}(t_{\rm c}) =  \left.\frac{\de V}{\de \bar{\phi}}\right|_{\bar{\phi}_{\rm c}},
\ee
such that for $ N < N_{\rm c}$, the dynamics is governed approximately by 
\begin{equation}
\ddot{\bar{\phi}} + 3H(t)\dot{ \bar{\phi}} = 0,
\label{eom-usr}
\end{equation}
and the Friedmann equation~\eqref{fried},
with $V_{\rm m}$ now specifying the plateau's height.
Fixing $(\bar{\phi}_{\rm m},\dot {\bar{\phi}}_{\rm m})$ at some point along this flat region, the system can be integrated exactly to yield
\be
\bar{\phi}(N) = \bar{\phi}_{\rm m} + \sqrt\frac23 M_{\rm Pl} \; \left[\tanh^{-1} \left(\sqrt{1+ \frac{2V_{\rm m}}{\dot {{\bar{\phi}}}^2_{\rm m} } }\right)- \tanh^{-1} \left(\sqrt{1+ \frac{2V_{\rm m}}{\dot {{\bar{\phi}}}^2_{\rm m} } e^{6(N-N_{\rm m})}}\right) \right]
\label{sol-usr}
\ee
and
\be
H^2(N)=\frac{1}{6M_{\rm Pl}^2} \left(\dot {{\bar{\phi}}}^2(N)  + 2V_{\rm m}\right), \quad\text{with}\quad \dot {{\bar{\phi}}}(N)=\dot {{\bar{\phi}}}_{\rm m} e^{-3(N-N_{\rm m})},
\label{fried-usr}
\ee
where we have switched to the e-fold time variable.

From the velocity in~\eqref{fried-usr}, we see that whatever value $\dot {{\bar{\phi}}}_{\rm m}$ has, the Hubble friction will (asymptotically) force the field to a halt~\cite{Tsamis:2003px}, say at position $\bar{\phi}_0$. The higher the initial velocity the longer it takes for this to happen. 
Moreover, from the Friedmann equation, this implies that the spacetime approaches dS even faster. 
To quantify this, let us parametrise the inflationary scale via the Lyth bound~\cite{Lyth:1996im} as $H \simeq 10^{-5} r M_{\rm Pl}$, with $r$ the tensor-to-scalar ratio, and assume that for most of the $60$ observable e-folds the field is slowly rolling down its potential. Since the potential will eventually dominate the inflationary motion, from~\eqref{fried-usr} we have
\be
V_{\rm m} = 10^{-10}r^2 M_{\rm Pl}^4.
\ee
Moreover, as already mentioned, the most conservative upper bound we can impose on the initial velocity is the Planck scale
\be
|\dot{{\bar{\phi}}}_{\rm m}| \leq M_{\rm Pl}^2 .
\ee

The interval $\Delta N_0\equiv |N_{\rm m} - N_0|$, during which the field decelerates while the Hubble rate approaches a constant, can be estimated by the condition
\be
\dot {\bar{\phi}}_{\rm m}^2 e^{-6\Delta N_{\rm 0}} = 2V_{\rm m},
\label{halt}
\ee
yielding
\be
\Delta N_{\rm 0} = \frac16 \ln \frac{\dot {\bar{\phi}}^2_{\rm m}}{2V_{\rm m}},
\label{Nds}
\ee
during which the classical solution~\eqref{sol-usr} will have travelled 
\be
\Delta \bar{\phi}_{\rm 0} \equiv|\bar{\phi}_0-\bar{\phi}_{\rm m}|\simeq \sqrt6\Delta N_{\rm 0} \;M_{\rm Pl}.
\label{phids}
\ee

For the extremal case $\dot{\bar{\phi}}_{\rm m}= M_{\rm Pl}^2$ and for $r\sim0.01$, from Eq.~\eqref{Nds} we obtain $\Delta N_{\rm 0} \simeq 4$ and $\Delta \bar{\phi}_{\rm 0}\sim 10 M_{\rm Pl}$. 
Alternatively, for $\dot {\bar{\phi}}_{\rm m}^2 < 2V_{\rm m}$, this happens within $\Delta N_{\rm 0}\lesssim 1$, while $\Delta \bar{\phi}_{\rm 0}\lesssim M_{\rm Pl}$. We thus see that such an extended plateau of at least $4$ e-folds is enough to dissipate away even a Planckian kinetic energy.
Due to the shift symmetry of the potential, we may now choose this terminal point\footnote{Note that the field only freezes asymptotically. The point defined as $\bar{\phi}_0$ is when the potential energy starts dominating the spacetime evolution.} as the origin of the inflationary trajectory, i.e. set $N_0=\bar{\phi}_0=0$, such that, for times $0<N<N_{\rm c}$, the solutions~\eqref{sol-usr},~\eqref{fried-usr} reduce to
\be
\bar{\phi}(N) \simeq 0  \quad\&\quad H^2 \simeq \frac{V_{\rm m}}{3M_{\rm Pl}^2}.
\label{cl_sol1}
\ee
As we will show in the next section, the quantum fluctuations then provide a natural initial condition for the slow-roll phase, which removes the fine-tuning of the initial phase-space to an adequate degree. 

However, as already mentioned, the price to pay for inserting such a decelerating plateau is that we have effectively turned a small-field model into a large-field one: even if during the $60$ e-folds of slow-roll inflation --- which begins at some time $N_{\rm i}\geq 0$ --- the field variation remains sub-Planckian (thus avoiding a large tensor background), we see that from $\phi_{\rm m}$ till $\phi_0$ the field can traverse a region of $10 M_{\rm Pl}$. Even though, due to the flatness of this region, the potential and kinetic energy differences remain always bounded by $M_{\rm Pl}$, a trans-Planckian field displacement can be problematic~\cite{Klaewer:2016kiy}. This is an ongoing issue affecting all large-field models which we will not address in this paper. 

\begin{figure}[h]
\centering
  \includegraphics[width=.8\linewidth]{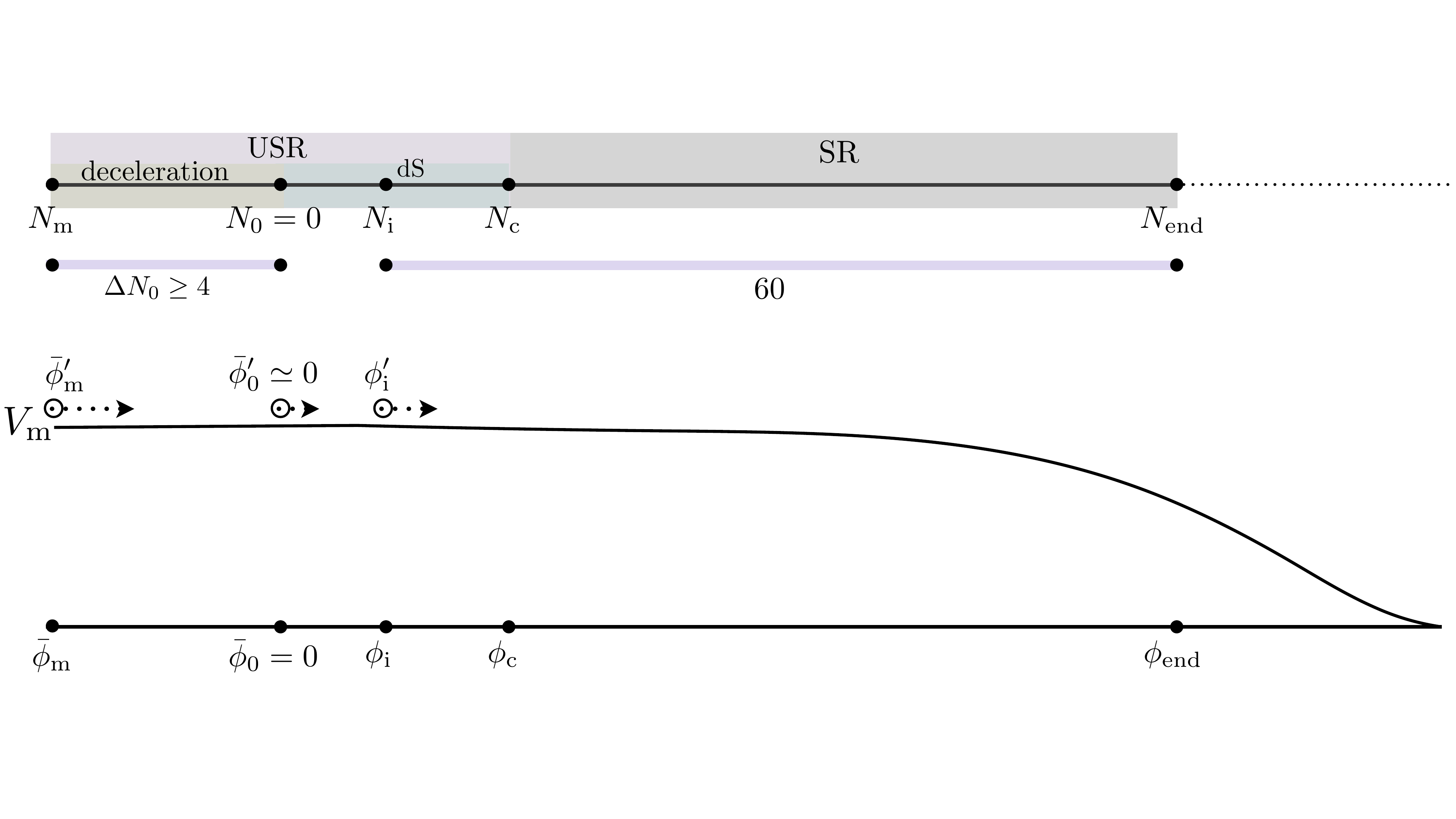}
 \caption{The relevant time-scales, the inflaton potential exhibiting a flat region and the corresponding field values and velocities.}
\label{fig:line}
\end{figure}

Before discussing the perturbations, let us summarise the time scales involved in the problem (see also Fig.~\ref{fig:line}), which are as follows:
\begin{itemize} 
\item[$N_{\rm m}$:] some arbitrary early time when the effective description of the system becomes operative. We take $\phi_{\rm m}$ lying along a potential plateau with some velocity $|\dot{\phi}_{\rm m}|\lesssim M_{\rm Pl}^2$ setting the UV cutoff
\item[$N_0$:] the time when the initial kinetic energy becomes smaller than the potential and the spacetime reaches dS. $\Delta N_{\rm 0}\equiv|N_{\rm 0}-N_{\rm m}|$ defines the interval during which the field is decelerating. Since a Planckian kinetic energy takes around $4$ e-folds to dissipate, in order to cover all phase-space we impose that this region last at least $4$ e-folds. Exploiting the shift symmetry along the nearly flat region we set $\phi_0=N_{\rm 0}=0$ 
\item[$N_{\rm c}$:] the time when the curvature of the potential becomes relevant marking the end of the USR phase. At this point the dynamics transits to slow-roll
\item[$N_{\rm i}$:] the time when the observable CMB scales leave the Hubble horizon. This is where we set the initial conditions for the following approximate 60 e-folds of inflation. $N_{\rm i}$ can, in principle, be anywhere after $N_0$ (see however the discussion at the end of Sec.~\ref{sec:ICs})
\end{itemize}

\subsection{Dynamics of fluctuations} \label{sec:fluct}
The long wavelength $k<aH$ modes of $\delta \phi$ and its conjugate momentum\footnote{On an FLRW background, the conjugate momentum is $a^3 \dot\phi $. However, the volume factor will not be relevant for our purposes.} $\delta\dot{\phi}$ can be treated as classical Gaussian random fields, which may be identified with their root-mean-square values:
\be \label{rms-def}
\delta \phi(t) \sim \delta\phi_{\rm rms}(t) \equiv \sqrt{\big\langle \delta\phi^2(t)\big\rangle} \qquad \& \qquad \delta \dot{\phi}(t) \sim \delta \dot{\phi}_{\rm rms}(t) \equiv \sqrt{\big\langle \delta\dot{\phi}^2(t)\big\rangle}.
\ee
As argued in the previous section, at any time $N_{\rm i}\in [0,N_{\rm c}]$, when $\bar{\phi}=\bar{\phi}_{\rm i}$, the background inflaton, that is, the collection of long wavelength Fourier modes of the field~\eqref{split}, is dominated by the fluctuations~\eqref{rms-def}:
\be \label{dfl}
\phi_{\rm i} = \delta\phi_{\rm rms}(t_{\rm i}) \quad\&\quad \dot{\phi}_{\rm i} = \delta\dot\phi_{\rm rms}(t_{\rm i}).
\ee
Therefore, if we can compute these we may set the field and velocity values at time $N_{\rm i}$. This is an initial condition in the sense that, given the presence of the plateau, any field value before $N_0=0$  is irrelevant; thus the interval $[0,N_{\rm c}]$ spans the initial patch of the phase-space. $N_{\rm i}$ can thus be thought of as just a coordinate charting the one-dimensional phase-space, such that for any $N_{\rm i}\in [0,N_{\rm c}]$ there is a fixed trajectory, which, as we show in the next section, does not overshoot. 

As already stressed, this rather trades the problem of a finely-tuned initial phase-space patch with that of a nearly flat region around the maximum, than solving it. Doing so though, allows one to argue that the latter situation can be justifiable on theoretical grounds. For example, in the context of the discussion around Eq.~\eqref{v-example}, one can ask what particle masses and couplings would be necessary in order to make inflation a generic outcome, i.e. in order to sustain at least $4$ e-folds of USR before the $60$ e-folds of slow-roll inflation.

In order to compute the rms values of \eqref{rms-def}, we may notice that around $N_0$, the fluctuation $\delta\phi(t_{\rm i},\x)$ and thus $\phi(t_{\rm i},\x)$ itself, can be treated as a quantum field on de Sitter space~\cite{Guth:1985ya} sourced by $V_{\rm m}$ as in~\eqref{cl_sol1}.
We may then consider the coincident limit of the 2-point field and velocity correlation functions at some time $t_0=0$ when the two fluctuations are in causal contact, as defining the background rms values at some later time $t_{\rm i}$, when the modes are well beyond the Hubble scale:
\be \label{coin-lim-def}
\phi_{\rm i} = \sqrt{\lim_{\x\to \y} \big\langle \phi(t_{\rm i},\x)\phi(t_{\rm i},\y)\big\rangle} \quad\& \quad \dot{\phi}_{\rm i} = \sqrt{\lim_{\x\to \y} \big\langle \dot{\phi}(t_{\rm i},\x)\dot{\phi}(t_{\rm i},\y)\big\rangle}.
\ee

We may now proceed with the canonical quantisation of a scalar tachyon on a fixed dS spacetime obeying 
\be
\ddot\phi +3H\dot\phi -\nabla^2\phi +\frac{\de V}{\de\phi}=0, 
\ee
with a potential of the form~\eqref{scalar_potential1}, in the standard way:
\begin{align}
&\phi(\tau,\x)
=\int\frac{\de^3 k}{(2\pi)^3}e^{i\K\x}\phi_{k}(\tau), \qquad
\phi_{k}(\tau) =
a(\K)v_k(\tau)+a^\dagger(-\K)v^*_k(\tau),
\end{align}
where the creation/annihilation operators satisfy canonical commutation relations: $[a(\K),a^\dagger(\q)]=(2\pi)^3\delta^3(\K-{\q})$, while the amplitude $v_k(\tau)$ obeying a Bunch-Davies asymptotic condition reads:
\begin{equation} \label{dS-mode}
v_k(\tau)=\frac{\sqrt{\pi}}{2}H(-\tau)^{3/2}{\rm H}^{(1)}_\nu(-k\tau).
\end{equation}
Here, ${\rm H}^{(1)}_\nu$ is the Hankel function of the first kind and we have defined
\begin{equation}
\nu \equiv \sqrt{\frac94 + \mu^2}\quad\text{with}\quad\mu^2\equiv \frac{m^2}{H^2},
\label{mu_and_nu}
\end{equation}
with the tachyon mass $m$ related to $|\eta|$ in~\eqref{scalar_potential1} as $m^2=6|\eta|H^2$.
The coincident limit correlators defining the rms values in~\eqref{coin-lim-def} now read 
\begin{equation} \label{2pt}
\big\langle \phi^2_{\rm i}\big\rangle
=\int \frac{\de^3 k}{(2\pi)^3} \left|v_k(\tau_{\rm i}) \right|^2,
\end{equation}
and
\begin{equation}
\label{dotphi2m0-def}
\big\langle \dot\phi^2_{\rm i}\big\rangle
=\int \frac{\de^3 k}{(2\pi)^3}H^2\tau_{\rm i}^2 \left| \frac{\de v_k}{\de\tau_{\rm i}} \right|^2.
\end{equation}

In the strictly massless limit, the field correlator is IR divergent~\cite{Vilenkin:1982wt,Linde:1982uu,Starobinsky:1982ee,Allen:1985ux,Ratra:1984yq, Ford:1984hs, Starobinsky:1986fx, Antoniadis:1986sb,Habib:1992ci}, while the velocity one is not. Indeed, for any non-vanishing velocity the field will reach an infinite value at $k\tau\to0$. For a non-vanishing mass, the massive and tachyonic cases are distinct. Had we been discussing a (non-tachyonic) light field, the Bessel index $\nu$ in Eq.~\eqref{mu_and_nu} would have been $\nu\to\bar\nu=3/2-\mu^2/3$ and for $\mu\neq0$ the integral \eqref{2pt} could have been evaluated exactly with the use of a UV cutoff. 
In other words, for a concave potential the field will travel only a finite distance before it rolls back down towards zero. In our case though, a tachyonic instability of the form \eqref{scalar_potential1}, renders both correlators IR divergent: a convex potential will eventually drive the field and its velocity to infinity.
Here, of course, the potential is tachyonic locally around the maximum and at some point higher order terms will regularise the correlator. Nevertheless, for our purposes, the dS horizon and the finite duration of the dS phase --- see Fig.~\ref{fig:line} --- provide a natural cutoff which regularises the divergence, even without considering higher terms. 

In order to see this, let us parametrise the physical momentum range as 
\be
\Lambda_{\rm IR} < \frac k{a(t)} <\Lambda_{\rm UV}.
\ee
Now consider initial (UV) and final (IR) physical time-slices marked by $t_0$ and $t_{\rm i}$ and a time $t\in [t_0,t_{\rm i}]$. These are to be identified with the corresponding times summarised in Fig.~\ref{fig:line}, so we can set $t_0=0$ and normalise the scale factor to $a_0=1$. The smallest upper bound for the co-moving momentum in the interval $[0,t_{\rm i}]$ is thus $k<\Lambda_{\rm UV}$, since $1<a<a_{\rm i}$. Similarly, the largest lower bound is given by $\Lambda_{\rm IR} a_{\rm i}<k$. Thus, the co-moving momentum cutoffs read
\begin{equation} \label{kL}
k_{\rm IR}=\Lambda_{\rm IR} a_{\rm i} \qquad\text{and} \qquad k_{\rm UV}=\Lambda_{\rm UV}.
\end{equation}

Let us further refine these. Assuming that the Hubble rate is constant for $t\in [0,t_{\rm i}]$, the size of the horizon is $\lambda_H\sim1/H$ throughout the whole evolution. The co-moving wavelength that reaches this scale at $t_{\rm i}$ is given by $\lambda_{\rm UV}= \lambda_H/a_{\rm i}$, while the one that is comparable to the horizon size at $t_0=0$ is $\lambda_{\rm IR}=a_{\rm i}\lambda_H$. Then, the IR cutoff should be set such that physical scales that are larger than the horizon during the entire interval $[0,t_{\rm i}]$ do not affect the Minkowski limit. That is, we exclude scales that never were in causal contact for $t<t_{\rm i}$. For the UV cutoff, the reasoning is similar; we are interested in computing the 2-point correlator at the final time-slice $t_{\rm i}$. The modes that contribute to the homogeneous background are super-horizon at this point. Therefore, $k_{\rm UV}$ should be chosen such that the physical scales which are always sub-horizon during $t\in [0,t_{\rm i}]$ do not contribute to the background at time $t_{\rm i}$. Therefore, 
\begin{equation} \label{Ph-cut}
\Lambda_{\rm IR}= \frac{H}{a_{\rm i}} \qquad\text{and} \qquad \Lambda_{\rm UV}=H a_{\rm i}.
\end{equation}
Plugging these into Eq.~\eqref{kL}, the comoving cutoffs take the form 
\begin{equation} \label{Com-cut}
k_{\rm IR}=H  \qquad\text{and} \qquad k_{\rm UV}=H a_{\rm i}.
\end{equation}

With proper cutoffs defined and the mode function \eqref{dS-mode} at hand, we may now compute the coincident limit 2-point field and velocity correlators, at time $t_{\rm i}$. In the massless limit --- we discuss finite (tachyonic) mass corrections right below --- the Hankel function acquires a simple form: $i\sqrt{\pi z^3}{\rm H}^{(1)}_{3/2}(z)=\sqrt2 e^{iz}(1-iz)$, leading to~\cite{Habib:1992ci}
\begin{equation} 
\label{2ptint}
\big\langle  \phi^2_{\rm i} \big\rangle 
=\frac{H^2}{8\pi}\int_{e^{-N_{\rm i}}}^{1} \de z \, z^2\left|{\rm H}^{(1)}_{3/2}(z)\right|^2 = \left(\frac{H}{2\pi}\right)^2\left( N_{\rm i} + \frac12 \left( 1 - e^{-2N_{\rm i}} \right) \right)
\end{equation}
and
\begin{equation}
\label{dotphi2m0}
\big\langle {\phi_{\rm i}'}^2 \big\rangle
= \frac{H^2}{8\pi}\int_{e^{-N_{\rm i}}}^{1} \de z \;z
\left| \frac{\de}{\de z}\left(z^{3/2}{\rm H}^{(1)}_{3/2}(z)\right)\right|^2= \frac14 \left(\frac{H}{2\pi}\right)^2 \left( 1-e^{-4N_{\rm i}} \right),
\end{equation}
where we have switched to the e-fold number as a time variable.

For $\mu=0$, the field correlator is indeed divergent in the IR (defined by $N_{\rm i}\to\infty$), while the velocity is regular. However, in our set-up the dS dynamics only holds as long as the potential is flat enough such that the USR condition~\eqref{usr-def} is met. Therefore, the time $t_{\rm c}$ --- defined in~\eqref{usr-sr} --- provides a maximal IR cutoff. Also note that since the cutoffs properly take into account the existence of the dS horizon, we need to keep the low energy cutoff in the $\big\langle {\phi'}^2\big\rangle$ correlator even if it is IR regular, which will also serve to regularise the aforementioned divergence due to the finite tachyonic mass corrections.

Equations~\eqref{2ptint} and \eqref{dotphi2m0} can be viewed as a parametric form of the phase-space curve given by
\begin{equation}
\label{ph-sp-curve}
x_{\rm i}^2
= \frac12 \left( 1-\sqrt{1-4q_{\rm i}^2} - \ln\sqrt{1-4q_{\rm i}^2} \right) ,
\end{equation}
where
\be  \label{rms-fields}
x_{\rm i} \equiv \frac{\sqrt{\left<  \phi^2_{\rm i} \right>} }{H/2\pi} \quad\text\&\quad q_{\rm i} \equiv  \frac{\sqrt{\left<  \phi'^2_{\rm i} \right>} }{H/2\pi},
\ee
with $0\leq q_{\rm i} \leq \frac12 \sqrt{1-e^{-4N_{\rm c}}} $. 

Before passing to a specific example, where we draw initial conditions from~\eqref{ph-sp-curve}, a few comments are in order. First, note that another way to obtain the qualitative form of the correlators is the stochastic formalism\footnote{Indeed, in the $N_{\rm i}\to0$ limit, the field $2$-point function~\eqref{2ptint} can be considered as a solution to a Langevin equation of the form $\dot\phi=n(t)$, where $n(t)$ is a unit-normalised Gaussian noise with variance $\sigma^2\propto t$.} introduced in~\cite{Starobinsky:1986fx}. This approach captures correctly the cutoff dependence of the $2$-point functions~\eqref{2ptint} and~\eqref{dotphi2m0} but disagrees in the numerical coefficients as a result of the different treatment of short modes~\cite{Habib:1992ci}. 

Second, in Ref.~\cite{Habib:1992ci}, the field and leading-order velocity correlators~\eqref{2ptint}, \eqref{dotphi2m0} were derived in the context of a scalar field on an expanding spacetime for a general UV cutoff.
Here, we have placed the computation in the context of initial conditions for inflation, via the argument presented in the previous section, while taking into account the physical meaning of the cutoffs and the tachyonic nature of the mass correction. By doing so one can then view the IR cutoff as a parameter running along the one-dimensional phase-space~\eqref{ph-sp-curve} setting the initial point of the observable inflationary trajectory. 

\begin{figure}[h]
\centering
  \includegraphics[width=0.45\linewidth]{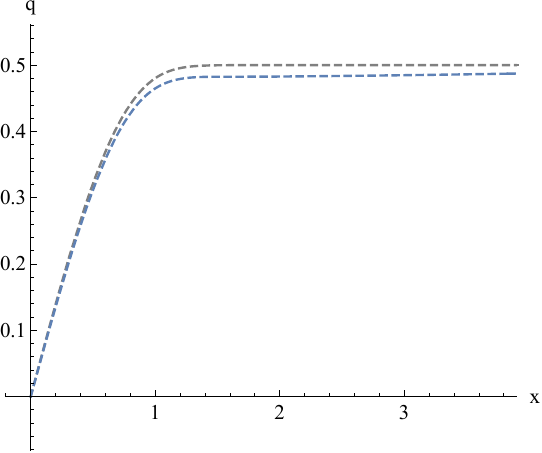}
 \caption{The phase-space deformation induced by considering $\mu^2 = 0.06$ (blue), compared to the $\mu^2 = 0$ case of Eq.~\eqref{ph-sp-curve} (gray). We take $0<N_{\rm i}<4$. }
\label{fig:phasespace_massive}
\end{figure}

Third, when the mass is non-vanishing, the time dependence of the field and velocity correlators, \eqref{2ptint} and \eqref{dotphi2m0}, gets modified and consequently the phase-space curve~\eqref{ph-sp-curve} gets deformed, albeit by a small amount. One may compute the corrections in $\mu$ order by order~\cite{Habib:1992ci} upon using known expressions for the derivatives of the Bessel functions with respect to the index. Alternatively, one can choose values for $\mu$ and $N_{\rm i}$ and perform the integrals numerically.
For example, the value of the mass that we will use in the next section in the context of hilltop inflation is $\mu^2=0.06$, while $N_{\rm i}$ runs from zero to its maximum value $N_{\rm c}$, with $N_{\rm c}=4$.
The corrections to the correlators in this case are less than $10\%$ of the leading $\mu=0$ values, which induces a negligible deformation of the phase-space curve as can be seen in Fig.~\ref{fig:phasespace_massive}. Nevertheless, in what follows we shall take the mass corrections into account by preforming the integrals \eqref{2pt} and \eqref{dotphi2m0-def} numerically.

Finally, note that our discussion is valid only when the field lies in the plateau region, i.e. for initial time $0<N_{\rm i}<N_{\rm c}$. Moreover, $N_{\rm i}$ must be small (order 1 number) since otherwise, as we comment in the next section, the scalar metric fluctuations induced by $\delta\phi$ dominate the background spacetime and one has to resort to a quantum treatment of gravitational dynamics. Therefore, our results hold as long as $N_{\rm i}$ (the time when CMB scales exit the horizon) is not far from $N_0$ (where the potential becomes the dominant contribution to the Hubble flow, which is where we set our zero). Thus the problem discussed here is the one of a fine-tuned initial kinetic energy. 

\section{Quadratic hilltop with quantum initial conditions}
\label{sec:hilltop}
As mentioned, our purpose is to argue that a USR plateau as a module of a small-field model can ease the initial condition problem by allowing sufficient time for the friction to slow down the background so that the quantum contributions~\eqref{2ptint} and~\eqref{dotphi2m0} become dominant.

Indeed, as argued around Eq.~\eqref{phids}, a plateau of at least $4$ e-folds (during which the field excursion is large) can dissipate away initial kinetic energies of the order of the Planck scale. 
As discussed in Sec.~\ref{sec:overshoot}, there are a few mechanisms that lead to such a flattened region away from the end of inflation, including interactions with massive fields, quantum corrections, etc. 

Let us note that in any case the USR dynamics must eventually come to an end giving way to the slow-roll phase~\cite{Namjoo:2012aa,Cai:2016ngx}. The reason is that during the USR plateau the Hubble slow-roll parameter is given by
\begin{equation}
\epsilon \equiv -\frac{\dot H}{H^2} = \frac{\dot\phi^2}{2H^2}\propto a^{-6}.
\label{e-usr}
\end{equation}
Upon defining the curvature perturbation in the comoving ($\delta\phi=0$) gauge via
\begin{equation}
\de s^2=-\de t^2 +a^2(t)e^{2{\cal R}(t,\x)}\de \x^2,
\label{R-def}
\end{equation}
we may compute its dimensionless power spectrum as
\begin{equation}
{\cal P}_\mathcal{R} = \frac{k^3}{2\pi^2} \left| {\cal R}_k\right|^2= \frac{H^2}{8\pi \epsilon},
\label{PS}
\end{equation}
and observe that due to Eq.~\eqref{e-usr} it receives an exponential enhancement of $e^{6N}$. This stems from the exponential growth of the would-be decaying super-horizon modes. Indeed, solving the Mukhanov-Sasaki equation~\cite{Sasaki:1983kd,Kodama:1985bj,Mukhanov:1988jd} on the USR background one finds on super-horizon scales
\be
{\cal R}=C_1+C_2\int \frac{\de t}{a^3\epsilon}.
\label{R}
\ee
In the slow-roll case where $\epsilon\ll1$ is constant, the $C_2$ mode would decay rapidly leaving the standard frozen contribution $C_1$ which sets the typical amplitude of a fluctuation. On the USR background however, from~\eqref{e-usr}, we see that this would-be decaying contribution now grows exponentially as $e^{3N}$, leading to a strong enhancement of the power spectrum. In order not to violate the CMB normalisation, we thus need to connect the USR initial phase to a slow-roll one; if that is not the case the growth during the $60$ e-folds would imply an inflationary energy scale so low that would have been already ruled out by surveys~\cite{Namjoo:2012aa,Cai:2016ngx}. 

Let us thus denote the end of the USR phase by $N_{\rm c}$, defined in~\eqref{usr-sr}, such that $\Delta N_{\rm c}\equiv N_{\rm c}-N_{\rm i}$ characterises how many of the observable 60 e-folds are spent in this phase. Since the modes freeze at the end of this period, during the transition to the slow-roll regime, the final value of the power spectrum reads
\begin{equation}
{\cal P}_\mathcal{R}= \frac{H^2}{8\pi \epsilon_{\rm c}}e^{6\Delta N_{\rm c}},
\label{P-usr}
\end{equation}
where $\epsilon_{\rm c}$ denotes the value of $\epsilon$ at the end of the USR phase. The observed value of the power spectrum is ${\cal P}_\mathcal{R} \approx 2.2 \times 10^{-9}$, while $\epsilon_c$ can be fixed by the tensor-to-scalar ratio $r$.  For $\Delta N_{\rm c} \sim 60$, (\ref{P-usr}) gives too small Hubble rate. The non-attractor USR inflation thus can only last for a limited number of $e$-folds~\cite{Namjoo:2012aa,Cai:2016ngx}: $\Delta N_{\rm c} \lesssim 4$.  
In order to study the transition to the slow-roll (SR) phase\footnote{The transition from the USR to the SR dynamics can also generate non-Gaussianity~\cite{Cai:2016ngx,Cai:2017bxr}.}, we may follow~\cite{Cai:2017bxr} and expand the potential around the transition point $\phi_{\rm c}$, where we have to match with observations, fixing $\epsilon_{\rm c}$ and $\eta_{\rm c}$.

In this note, however, we will assume the existence of a plateau smoothly extending the maximum at the hilltop into a region where~\eqref{usr-def} holds. Such a region may be the result of a running mass due to quantum corrections or of the higher order terms that stabilise the potential far from the maximum. We may now observe that hilltop inflation, by construction, begins in such a USR phase, i.e., there always exists a finite region around the maximum for which the condition~\eqref{usr-def} is satisfied; this region, however, is too short to serve as a decelerating patch. Therefore, we may set the point where the plateau is ``glued" to the hilltop to be $\phi_0=0$ and expand the potential around there. In other words, whatever the reason of an extended plateau around the maximum is, at $\phi=0$ we consider that the theory is well described by the standard quadratic tachyonic potential:
\begin{equation}
V(\phi) = 3 H^2 M_{\rm Pl}^2 \left(1 - \alpha ^2 \frac{ \phi}{M_{\rm Pl}} - \frac{\mu ^2}{6}\frac{\phi ^2}{M_{\rm Pl}^2} + \dots \right),
\label{scalar_potential2}
\end{equation}
where we have assumed a small but non-vanishing gradient. The dimensionless parameters $\alpha$ and $\mu$ are related to the slow-roll parameters at the origin as
\begin{eqnarray}
\epsilon_{_V } &\equiv& \frac{M_{\rm Pl}^2}{2}\left( \frac{V,_\phi}{V} \right)^2 = \frac{\alpha ^4}{2} ,  \\
\eta_{_V} &\equiv&  M_{\rm Pl}^2 \frac{V,_{\phi\phi}}{V}  = -\frac{\mu^2}{3} .
\label{slow-roll-parameter}
\end{eqnarray}

Around the maximum of the scalar potential, where $V(\phi)$ is given by (\ref{scalar_potential2}), the Klein-Gordon equation can be solved exactly to give
\begin{eqnarray}
\phi (N) =  e^{-\frac{3}{2}  (N-N_{\rm i})} \left(3M_{\rm Pl}\frac{\alpha ^2}{\mu ^2}+\phi_{\rm i}\right) \cosh \left[\nu  (N-N_{\rm i})\right]-3M_{\rm Pl}\frac{\alpha ^2}{\mu ^2}\nonumber\\
+ \frac{e^{-\frac{3}{2} (N-N_{\rm i})}}{2\nu/3 }  \left(3M_{\rm Pl}\frac{\alpha ^2}{\mu ^2} + \phi_{\rm i} + \frac23 \phi_{\rm i}^\prime \right)\sinh \left[\nu  (N-N_{\rm i})\right].
\label{exact_sol}
\end{eqnarray}
This solution satisfies the initial condition $\phi(N_{\rm i}) = \phi_{\rm i}$ and $\phi^\prime(N_{\rm i}) = \phi_{\rm i}^\prime$, with  $(\phi_{\rm i},\phi_{\rm i}')$ given by \eqref{2ptint} and \eqref{dotphi2m0}. Note that the classical solution is now uniquely determined by a single parameter, the initial time $N_{\rm i}$ charting the curve~\eqref{ph-sp-curve}. Let us define the ratio $\beta(N) \equiv \left|\frac{\partial V/\partial \phi} {\ddot\phi (t)}\right|$, such that the USR condition~\eqref{usr-def} is satisfied for $\beta \ll 1 $.  
As the field evolves, the curvature of the potential becomes more important with the transition from ultra slow-roll to slow-roll happening when $\beta(N_{\rm c}) = 1$. Upon using the above classical solution, the transition time $N_{\rm c}$ can be determined analytically:
\be
\Delta N_{\rm c} \equiv N_{\rm c} - N_{\rm i}  =\frac{1}{2\nu}  \log \left[\frac{\phi_{\rm i}^\prime \left(\nu +3/2\right) - 3 \alpha ^2 M_{\rm Pl} - \mu ^2 \phi_{\rm i}  }{\phi_{\rm i}^\prime \left(\nu -3/2\right) + 3 \alpha ^2 M_{\rm Pl} + \mu ^2 \phi_{\rm i} }\right].
\label{USR_DN}
\ee
\begin{figure}
\centering
\begin{subfigure}{.5\textwidth}
  \centering
  \includegraphics[width=1\linewidth]{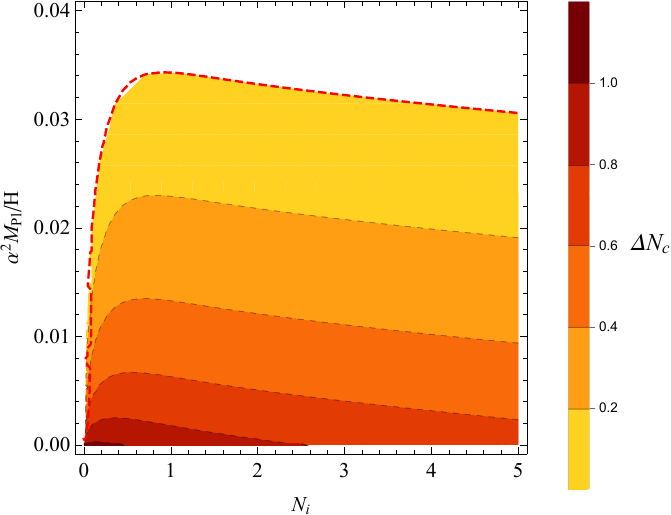}
  \label{fig:sub1}
\end{subfigure}%
\begin{subfigure}{.5\textwidth}
  \centering
  \includegraphics[width=1\linewidth]{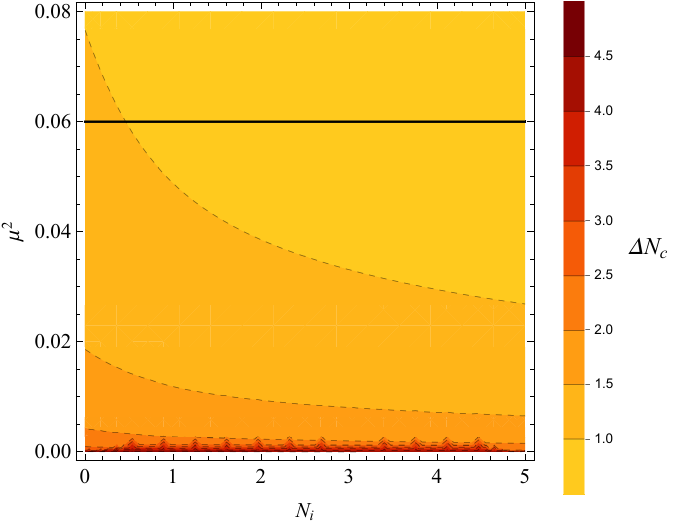}
  \label{fig:sub2}
\end{subfigure}
\caption{Left panel: contour plot for $\Delta N_{\rm c}$ showing the range of initial time $N_{\rm i}$ and slope of the potential $\alpha$ for which $\Delta N_{\rm c}>0$, with $\mu^2 = 0.06$. The red-dashed line corresponds to $\Delta N_{\rm c} = 0$, i.e. the initial time is taken in the slow-roll regime. Right panel: contour plot for $\Delta N_{\rm c}$ showing the range of initial time $N_{\rm i}$ and the parameter $\mu^2$ satisfying $\Delta N_{\rm c} > 0$. We consider the limit where $r = 0$. The black-solid line corresponds to $\mu^2 = 0.06$ (or $\eta_V = -0.02$). }
\label{fig:contours}
\end{figure}

Since our quantum computation holds for $0\leq N_{\rm i}\leq N_{\rm c}$, we have to impose $\Delta N_{\rm c} > 0$. In Fig.~\ref{fig:contours} (left panel) we show the allowed values for $N_{\rm i}$ and $\alpha$ that meet this condition accommodating a possibly observed USR phase for fixed $\mu^2 = 0.06$ (or equivalently $\eta_{_V} \approx - 0.02$). For each pair of points $(N_{\rm i}, \alpha^2\frac{M_{\rm Pl}}{H})$, the shading corresponds to the value of $\Delta N_{\rm c}$ obtained by (\ref{USR_DN}). As shown there, in order to have $\Delta N_{\rm c} > 0$, the slope $\alpha$ needs to satisfy the constraint
\begin{equation}
\alpha^2 \frac{M_{\rm Pl}}{H} \lesssim 0.0346.
\end{equation}
By using (\ref{slow-roll-parameter}) and $r = 16 \epsilon_V$, the existence of an observable USR phase gives the following constraint on the tensor-to-scalar ratio:
\begin{equation}
r \lesssim 10^{-2} \left(\frac{H}{M_{\rm Pl}} \right)^2.
\label{r-constraint}
\end{equation}
Hence, for $H$ around the GUT scale, the tensor-to-scalar ratio must have a vanishingly small value.  
We may thus consider the parameter space $(N_{\rm i}, \mu^2)$ in the $r = 0$ limit. We show the allowed values in the right panel of Fig.~\ref{fig:contours}. The shading corresponds to the value of $\Delta N_{\rm c}$ obtained by (\ref{USR_DN}). The maximum duration of the USR phase for $\mu^2 = 0.06$ can be determined by taking the limit $N_{\rm i} \rightarrow 0$ in which we get $\Delta N_{\rm c}^{\rm max} \approx 1$. 

\begin{figure}[ht]
\centering
  \includegraphics[width=0.45\linewidth]{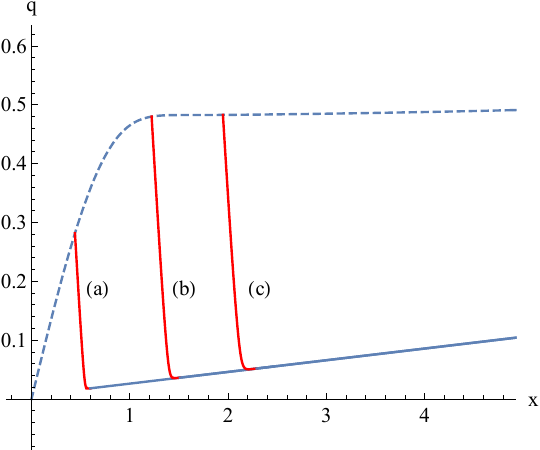}
 \caption{The blue dashed line represents the one-dimensional initial condition phase-space \eqref{ph-sp-curve} with finite mass corrections taken into account. Trajectories (a), (b), (c) correspond to the inflaton dynamics for an initial time $N_{\rm i} = (0.1,\; 1,\; 3),$ respectively. They start by following non-attractor trajectories and then transit to the slow-roll profile (blue solid line), which spans $\sim60$ e-folds. The red solid lines depict the first $4$ e-folds of each trajectory. We have fixed $\alpha = 0$ and $\mu^2 = 0.06$.}
\label{phasespace_Ni}
\end{figure}

In order to study the dynamics, we may introduce phase-space variables
\be  \label{phase-space-variable}
x(N) \equiv \frac{\phi(N)}{H/2\pi} \quad\text\&\quad q(N) \equiv  \frac{\phi^\prime(N) }{H/2\pi},
\ee
such that $x(N_{\rm i}) = x_{\rm i}$ and $q(N_{\rm i}) = q_{\rm i}$, with the latter defined in (\ref{rms-fields}).
In Fig.~\ref{phasespace_Ni}, we plot the solution of the equation of motion with initial conditions $(x_{\rm i},q_{\rm i})$ lying along the (mass corrected) curve~\eqref{ph-sp-curve} as dictated by the quantum fluctuations. As can be seen, the phase-space trajectories begin in the non-attractor phase and rapidly (within 4 e-folds) transit to the slow-roll profile $q \propto |\eta_{_V} |x$, which lasts for 60 e-folds, avoiding the overshoot issue~\cite{Bird:2008cp} and hence desensitising hilltop inflation from the initial kinetic energy. 

\section{Conclusions}\label{sec:conc}
Hilltop models are attractive from an effective field theory perspective but suffer from fine-tuning; for a successful (60 e-folds long) phase of slow-roll inflation, one has to begin inside a very small patch of the available phase-space. The reason why this happens is easy to understand: in small-field models there is not enough time for the Hubble friction to drive the dynamics towards the slow-roll attractor, hence one has to set initial conditions very close to it in order for this to happen within 60 e-folds.

In this work, we addressed the question of how one can relax this problem from a bottom-up perspective. To this end, we studied two solutions: one based on symmetries, where the inflaton starts its slow-rolling at the hilltop due to some putative internal symmetry which is subsequently spontaneously broken; and another, encompassing an ultra flat region around where one sets initial conditions as a generic module of any small-field model. In the second case, if the plateau is long enough, the classical field comes to a halt exponentially fast at which point the super-horizon quantum fluctuations dominate the background dynamics. The latter can be treated as quantum fields on de Sitter space whose correlators can be computed via the standard quantisation procedure. If CMB scales exit the Hubble horizon shortly after the classical background has frozen, the rms values of the quantum fluctuations provide initial conditions for the field and its velocity in the sense that whatever their values were before that becomes dynamically irrelevant.

Setting initial conditions in this way reduces the phase-space to one dimension and partially evades the overshoot problem in the sense that once the initial position is chosen close to the would-be maximum, the velocity is automatically fixed.
A plateau of around $4$ e-folds is enough to dissipate away even a Planckian initial kinetic energy at the cost of a trans-Planckian field displacement during this phase (although this takes place before the CMB scales exit the horizon). 
We have thus recasted the initial velocity problem in small-field models as a condition on the flatness of the potential around the initial field value by considering a ``large-field" module, which is well-motivated in embeddings of inflation in UV frameworks featuring a variety of additional degrees of freedom.

\section*{Acknowledgements}
This work was supported in part by a CNRS PICS Grant No. 07964 and in part by the ``CUniverse'' research promotion project by Chulalongkorn University (Grant Reference CUAASC) and the CU Global Partnership Programme (Grant No. B16F630071).

\end{document}